\documentclass[10pt,aps,twocolumn,floats,floatfix,amssymb,prd,superscriptaddress, 
noeprint]{revtex4-2}

\usepackage{graphicx}
\usepackage{amsmath,amssymb}
\usepackage{amsfonts}
\usepackage{xspace} 
\usepackage{xcolor}
\usepackage{dcolumn}
\usepackage{bm}
\usepackage{float}
\usepackage{footnote}
\usepackage{fontawesome}
\usepackage{mathrsfs}
\definecolor{Burgundy}{RGB}{144,0,32}
\usepackage[colorlinks=true,linkcolor=blue,citecolor=blue, filecolor=blue, urlcolor=blue]{hyperref}
\usepackage[all]{hypcap} 
\usepackage[utf8]{inputenc} 
\usepackage{multirow}
\usepackage{etoolbox}
\usepackage{tikz}
\usepackage{dcolumn}
\usepackage{array}
\usepackage[subnum]{cases}

\usepackage[caption=false]{subfig}

\begin{document}

\title{
Ringing of rapidly rotating black holes in effective field theory }

\author{Tom van der Steen}
\email{tom.vandersteen@kuleuven.be}
\affiliation{Institute for Theoretical Physics, KU Leuven,
Celestijnenlaan 200D, 3001 Leuven, Belgium}
\affiliation{Leuven Gravity Institute, KU Leuven,
Celestijnenlaan 200D, 3001 Leuven, Belgium}

\author{Simon Maenaut}
\email{simon.maenaut@nbi.ku.dk}
\affiliation{Center of Gravity, Niels Bohr Institute,
Blegdamsvej 17, 2100 Copenhagen, Denmark}

\author{Stef~J.~B.~Husken}
\email{stef.husken@student.kuleuven.be}
\affiliation{Institute for Theoretical Physics, KU Leuven,
Celestijnenlaan 200D, 3001 Leuven, Belgium}
\affiliation{Leuven Gravity Institute, KU Leuven,
Celestijnenlaan 200D, 3001 Leuven, Belgium}

\author{Pedro~G.~S.~Fernandes}
\affiliation{Institut f\"ur Theoretische Physik, Universit\"at Heidelberg, Philosophenweg 12 \& 16, 69120 Heidelberg, Germany}

\author{Maxim~D.~Jockwer}
\affiliation{Institute for Theoretical Physics, KU Leuven,
Celestijnenlaan 200D, 3001 Leuven, Belgium}

\author{Vitor~Cardoso}
\affiliation{Center of Gravity, Niels Bohr Institute, Blegdamsvej 17, 2100 Copenhagen, Denmark}
\affiliation{CENTRA, Departamento de Física, Instituto Superior Técnico – IST,
Universidade de Lisboa – UL, Avenida Rovisco Pais 1, 1049-001 Lisboa, Portugal}

\author{Thomas~Hertog}
\affiliation{Institute for Theoretical Physics, KU Leuven,
Celestijnenlaan 200D, 3001 Leuven, Belgium}
\affiliation{Leuven Gravity Institute, KU Leuven,
Celestijnenlaan 200D, 3001 Leuven, Belgium}

\author{Tjonnie~G.~F.~Li}
\affiliation{Department of Physics and Astronomy, Laboratory for Semiconductor Physics, KU Leuven, B-3001 Leuven, Belgium}
\affiliation{Leuven Gravity Institute, KU Leuven,
Celestijnenlaan 200D, 3001 Leuven, Belgium}
\affiliation{STADIUS Center for Dynamical Systems, Signal Processing and Data Analytics,
KU Leuven, Kasteelpark Arenberg 10, 3001 Leuven, Belgium}

\begin{abstract}

Within the effective field theory approach to gravity, deviations from general relativity can be systematically described by higher-curvature operators. However, computing the resulting corrections to black hole quasinormal mode spectra remains challenging in the rapidly rotating regime, where perturbative expansions in the spin break down.
We use recently constructed numerical rotating black hole solutions to compute quasinormal mode frequency corrections at leading order in the effective field theory. Focusing on scalar perturbations, we evaluate cubic-curvature corrections, which constitute the leading modifications.
We employ a pseudo-spectral collocation method to solve the resulting perturbation equations on these backgrounds, enabling accurate computation across a broad parameter range. We obtain frequency corrections for fundamental modes with $l\le5$ for all $m$, and the first overtone of $2 \le l \le 5$ modes for all $m$ for spins up to $a=0.99M$, with relative errors below $10^{-4}$.
We observe that corrections to certain modes grow significantly as the spin approaches the near-extremal regime.

\end{abstract}

\maketitle

\section{Introduction}

Gravitational waves (GWs) have opened a whole new window into the universe to study very massive compact objects and the nature of gravity itself \cite{Cardoso:2016ryw,bertiExtremeGravityTests2018a,bertiExtremeGravityTests2018b,cardosoGravitationalwaveSignaturesExotic2016,askarBlackHolesGravitational2019,Cardoso:2019rvt}. They carry detailed information about gravity in a highly warped and dynamical regime and directly probe black holes (BHs), one of general relativity's most remarkable predictions. As the number of observations of BH binary mergers and their precision increases \cite{LVKcollaborationGWTC40UpdatingGravitationalWave2025, LVKcollaborationGWTC3CompactBinary2023}, we are able to perform more precise tests of General Relativity (GR) and constrain any deviations from it \cite{LVKcollaborationGW250114TestingHawkings2025, LVKcollaborationTestsGeneralRelativity2025,LVKcollaborationTestsGeneralRelativity2026a,LVKcollaborationTestsGeneralRelativity2026b,LVKcollaborationTestsGeneralRelativity2026c}. 

The final stage of a BH binary coalescence is particularly well-suited to perform such tests of gravity. In the ringdown phase, the newly-formed BH quickly relaxes to a stationary state. The dynamics of this perturbed BH are governed by its quasinormal modes (QNMs), a discrete set of characteristic damped oscillations \cite{bertiGravitationalwaveSpectroscopyMassive2006, bertiQuasinormalModesBlack2009, bertiBlackHoleSpectroscopy2025, vishveshwaraStabilitySchwarzschildMetric1970, teukolskyRotatingBlackHoles1972}. The complex frequencies of these QNMs are fully determined by the background geometry, and consequently provide a clean probe of the remnant BH.

Within GR, stationary BHs can be fully characterized by only their mass, spin and charge, a result known as the no-hair theorem. Since astrophysical BHs have negligible charge, the Kerr metric provides a full description of these BHs. Therefore the characteristic QNM frequencies also only depend on the properties of the BH. Hence, verifying the consistency in the inferred mass and spin from multiple QNMs provides a strong test of GR. The study of BH spectra is also known as BH spectroscopy \cite{dreyerBlackholeSpectroscopyTesting2004, Berti:2005ys,bertiSpectroscopyKerrBlack2016, carulloEmpiricalTestsBlack2018, isiTestingNohairTheorem2019, franchiniTestingGeneralRelativity2024, ligoscientificGW250114TestingHawkings2025, LVKcollaborationBlackHoleSpectroscopy2026}.
In the simplest case, this provides a clear null test of GR: any inconsistencies in the inferred parameters would signal that the no-hair theorem is violated, although this would not directly provide an indication for the possible nature of such deviations from GR.

To interpret potential deviations in the QNM spectrum requires concrete theoretical predictions in theories beyond GR. To navigate the vast landscape of theories, it is natural to adopt an effective field theory (EFT) approach to gravity, in which the Einstein–Hilbert action is supplemented by higher-derivative operators suppressed by a high energy scale \cite{stelleRenormalizationHigherderivativeQuantum1977, stelleClassicalGravityHigher1978, donoghueGeneralRelativityEffective1994}. This framework provides a general and systematic treatment of extensions of GR, also known as higher-derivative theories, that are consistent with its empirical success across scales \cite{willConfrontationGeneralRelativity2014, yunesGravitationalWaveTestsGeneral2024}, and respect its symmetries. We further restrict to theories that do not introduce new light degrees of freedom. The Lorentz and diffeomorphism invariance of GR require that the higher-derivative operators enter the expansion as scalar quantities, and fixing the light degrees of freedom imposes that any modifications can only depend on the metric. The result is a unique expansion of the action in curvature invariants \cite{endlichEffectiveFormalismTesting2017, cardosoBlackHolesEffective2018}. Any theory satisfying these assumptions can be mapped onto a small set of observational constraints on these higher-curvature couplings \cite{maenautRingdownAnalysisRotating2026}.

In this EFT expansion of GR, the leading, purely-gravitational corrections to vacuum BH solutions arise at cubic order, unlike other higher-curvature theories. In scalar–tensor theories that include additional scalar degrees of freedom, corrections appear at quadratic order in curvature already, since the scalars dynamically couple to the quadratic curvature invariants \cite{horndeskiSecondorderScalartensorField1974, yunesDynamicalChernSimonsModified2009,sotiriouBlackHoleHair2014}. BHs in scalar-tensor theories admit scalar hair, and lead to a separate set of constraints \cite{chungProbingQuadraticGravity2025}. 

There has been an ongoing effort to quantify QNM frequency corrections in the EFT framework. However, it has proven difficult to extend this to rapidly spinning BHs, which are most relevant for observations \cite{loustoRemnantMassesSpins2010, LVKcollaborationGWTC40UpdatingGravitationalWave2025}. In theories beyond GR, the computation of QNMs is complicated by modified background geometries, altered boundary conditions, and perturbation equations that are generally no longer separable \cite{franchiniTestingGeneralRelativity2024, konoplyaQuasinormalModesBlack2011}. Within the EFT framework, initial results were obtained for non-rotating BHs \cite{cardosoPerturbationsSchwarzschildBlakcholes2009, molinaGravitationalSignatureSchwarzschild2010, blazquezsalcedoPerturbedBlackHoles2016, blazquezsalcedoQuasinormalmodesEinsteingaussbonnetdilatonBlackholes2017, tattersallQuasinormalmodesBlackholesHorndeski2018, cardosoBlackHolesEffective2018, cardosoParametrizedBlackholeQuasinormal2019, mcmanusParametrizedBlackholeQuasinormal2019, konoplyaQuasinormalModesStability2020, derhamBlackholeGravitationalWaves2020, silvaQuasinormalModesExcitation2024, alyModifiedTeukolskyFormalism2026}, and later extended to linear and quadratic order in the spin \cite{pieriniQuasinormalmodesRotatingBlackholes2021, srivastavaAnalyticalComputationQuasinormalmodes2021, canoGravitationalRingingRotating2022, pratikQuasinormalModesSlowlyrotating2022, pieriniQuasinormalmodesRotatingBlackholes2022}. The background solution for moderately spinning BHs was obtained, using a spin expansion that could, in principle, go to arbitrary order \cite{canoLeadingHigherderivativeCorrections2019,canoAccuracySlowrotationApproximation2024}. Using this analytical background, the corrections to the QNMs could be accurately determined up to dimensionless spins of $a/M \lesssim 0.7$ \cite{canoRingingRotatingBlack2020, canoUniversalTeukolskyEquations2023, canoQuasinormalModesRotating2023, canoParametrizedQuasinormalMode2024, canoHigherderivativeCorrectionsKerr2024, chungQuasinormalModeFrequencies2024, chungQuasinormalModeFrequencies2025}, or alternatively using spectral methods up to $a/M \lesssim 0.8$ \cite{blazquez-salcedoQuasinormalModeSpectrum2025, blazquez-salcedoQuasinormalModesRapidly2025a, khooQuasinormalModesRotating2025}. However, a systematic and full characterization of QNM spectra in the EFT framework at high spins remained out of reach.

In this work, we extend the corrections to the spectrum to rapidly spinning BHs. Recent advances, using spectral methods, have enabled the construction of highly accurate rotating BH backgrounds in higher-derivative gravity, even for high spins \cite{fernandesLeadingEffectiveField2025, lamSpinningBlackHoles2026, lamAnalyticAccurateApproximate2026}. Building on these developments, we compute the corrections to the scalar QNM frequencies, using pseudo-spectral collocation methods on the new numerical BH backgrounds. These scalar perturbations are sensitive to higher-curvature corrections in the background geometry, and can be used as a proxy for the gravitational QNMs \cite{paniScalarShortcutBeyondKerr2026}. We focus on the leading, non-trivial corrections at cubic order within the EFT framework. We provide a systematic characterization of scalar perturbations across large spins by computing QNM frequencies for multiple fundamental modes with $l \leq 5$ and corresponding $m$ up to $a/M\le0.99$, as well as the first overtone with $l=m=2$ up to $a/M\le 0.99$. Our results provide the first systematic, high-spin characterization of scalar QNM spectra in the EFT framework to gravity (see also \cite{huskenQuadraticGravityCorrections2026, miguelEFTCorrectionsScalar2024} for scalar-tensor theories).

Very recently, corrections to the dominant $(l,m,n)=(2,2,0)$ gravitational mode in the EFT approach for gravity were obtained at high spins in \cite{boyceKerrBlackHole2026}. While this presents an important first step, a systematic exploration of the spectrum across different harmonics and overtones remains largely unexplored. This limits the application to precision ringdown tests beyond GR within an EFT framework. It was shown in \cite{paniScalarShortcutBeyondKerr2026} that scalar QNMs can serve as a reliable proxy for gravitational QNMs in beyond-GR scenarios, accurately capturing their deviations from GR. Leveraging this correspondence, our results provide a complementary and extensible approach to perform ringdown tests in the EFT framework.

Our paper is set up as follows: in Sec.~\ref{sec:cubic} we lay out how the Klein-Gordon equation on the modified BH backgrounds can be expanded to leading order. 
In Sec.~\ref{sec:spectral} we elaborate on the numerical collocation method used to solve these equations. In Sec.~\ref{sec:results} we summarize the results and in Sec.~\ref{sec:discussion} we discuss the connection between our results and the Eikonal limit.

\section{Scalar QNMs in higher-derivative gravity}\label{sec:cubic} 
We consider higher-derivative corrections to the Einstein-Hilbert action. Including the leading-order, purely gravitational corrections that contribute non-trivially to a BH background in four dimensions, the action is (up to field redefinitions) given by \cite{canoLeadingHigherderivativeCorrections2019}: 
\begin{align}
    S=&\frac{1}{16\pi}\int d^4x \sqrt{-g}\Big[R
    +\lambda_\mathrm{ev}\ell^4R_{\mu\nu}{}^{\rho\sigma}R_{\rho\sigma}{}^{\delta\gamma}R_{\delta\gamma}{}^{\mu\nu}\Big],
    \label{eq:action}
\end{align}
where $\ell$ is the EFT length scale suppressing higher-curvature terms and $\lambda_\mathrm{ev}$ is a dimensionless coupling constant. We will only focus on the parity-even corrections, since the parity-odd corrections do not modify the scalar QNMs \cite{canoRingingRotatingBlack2020}. We have also confirmed, using our numerical method, that these modifications do indeed vanish. The physical curvature scale of these systems is set by their mass $M$. Hence, we can introduce a dimensionless coefficient $\lambda=\lambda_\mathrm{ev}\ell^4/M^4$, which characterizes the size of the corrections compared to GR. We work perturbatively in $\lambda$ and assume that $|\lambda|\ll1$. 

In higher-derivative gravity, the Kerr solution is no longer a solution of the modified equations of motion (EOM). Nevertheless, since the higher-derivative coupling $\lambda$ is small, we can still find rotating BH solutions in this modified theory by expanding both the EOM and the metric around the GR solution in powers of $\lambda$. For a rotating black hole with mass $M$ and spin $a$, this yields the following corrected Kerr geometry \cite{canoLeadingHigherderivativeCorrections2019}:
\begin{equation}\label{eq:background}
\begin{split}
	g_{\mu\nu}dx^\mu dx^\nu =&
    -\left(1-\frac{2Mr}{\Sigma}-\lambda H_1\right) dt^2 \\
	& - (1+\lambda H_2)\frac{4 M ar (1-y^2)}{\Sigma} dt d\phi\\
	&+(1+\lambda H_3)\Sigma\left(\frac{dr^2}{\Delta}+\frac{dy^2}{1-y^2}\right)\\
	&+(1+\lambda H_4)\left(r^2+a^2+\frac{2a^2Mr(1-y^2)}{\Sigma}\right)\\
    &\times(1-y^2)d\phi^2,
\end{split}
\end{equation}
in Boyer-Lindquist coordinates $x^\mu=(t,r,\phi,y)$ with $y=\cos(\theta)$. Here, we have also introduced the quantities
\begin{equation}
	\Delta = r^2 - 2Mr +a^2, \quad \Sigma = r^2 + a^2 y^2.
\end{equation}
We encode the metric corrections as the functions $H_i(r,y)$, that solely depend on the radial and polar coordinate. In this parameterisation, the locations of the horizons remain at the roots of $\Delta$, namely $r_\pm=M\pm\sqrt{M^2-a^2}$, to linear order in $\lambda$. By imposing the right asymptotic behaviour (see Eq.~(3.11) of \cite{canoLeadingHigherderivativeCorrections2019}), the meaning of $M$ and $a$ being the ADM mass and spin of the solution is preserved.

Scalar QNMs are solutions of the massless Klein–Gordon equation on the BH background that satisfy quasinormal, dissipative boundary conditions: purely ingoing at the event horizon and purely outgoing at infinity. Hence, we are looking for solutions $\psi$ to the Klein-Gordon equation on a curved background
\begin{equation}\label{eq:KleinGordon}
    \Box\psi = \frac{1}{\sqrt{-g}}\partial_\mu(g^{\mu\nu}\sqrt{-g}\partial_\nu\psi) = 0,
\end{equation}
that satisfy the proper boundary conditions. First of all, we observe that, due to the background being a stationary, axisymmetric solution, the PDE in Eq.~\eqref{eq:KleinGordon} can be separated along the Killing directions $\partial_t$ and $\partial_\phi$ by taking
\begin{equation}
    \psi(t,r,y,\phi)=\Phi(t,\phi)\psi_{m,\omega}(r,y)
\end{equation}
where $\Phi(t,\phi)=e^{-i(\omega t-m\phi)}$ for some frequency $\omega\in\mathbb{C}$ and azimuthal number $m\in\mathbb{Z}$ \cite{teukolskyPerturbationsRotatingBlack1973, miguelEFTCorrectionsScalar2024}. This way, we can introduce a new operator that only acts on the $(r,y)$-coordinates:
\begin{equation}\label{eq:D_m_omega}
    \mathcal{D}_{m}(\omega)[\psi_{m,\omega}]\equiv \frac{1}{\Phi}\Box\psi=0.
\end{equation}
For a fixed azimuthal number $m$, the QNM frequencies $\omega$ are the complex values for which the operator $\mathcal{D}_{m}(\omega)$ admits a non-trivial solution satisfying the QNM boundary conditions. This defines a linear, second order boundary value problem (BVP) in the coordinates $(r,y)$. 

Before we address the boundary conditions (BCs) for $\psi_{m,\omega}$, we first carry out the perturbative expansion in $\lambda$ of Eq.~\eqref{eq:D_m_omega}. Since the background in Eq.~\eqref{eq:background}  is perturbatively close to the Kerr solution, the Klein-Gordon operator $\mathcal{D}_{m}(\omega)$ splits up in an operator for the Kerr background plus one governing the linear corrections. These operators take the form \cite{canoRingingRotatingBlack2020}:
\begin{align}
    \mathcal{D}^{(0)} \psi &= 
	\frac{1}{\Sigma} \partial_r (\Delta \partial_r \psi) 
	+\frac{1}{\Sigma} \partial_y \left((1-y^2)\partial_y \psi\right)\notag\\
	&-\frac{\psi(1-y^2)}{\Delta \Sigma} \bigg[
	\frac{(\Sigma-2Mr)m^2}{(1-y^2)^2}
	+ \frac{4M ar m \omega}{1-y^2}\notag\\
	&-\left(2Mr a^2 + \frac{\Sigma(r^2+a^2)}{1-y^2}\right)\omega^2
	\bigg],\label{eq:D0}\\
    \mathcal{D}^{(1)} \psi &= 
	-\frac{H_3\Delta}{\Sigma}\partial_r^2 \psi 
	+\frac{H_3(y^2-1)}{\Sigma}\partial_y^2 \psi \notag\\
	&+ \frac{P}{\Sigma} \partial_r \psi 
	+ \frac{Q}{\Sigma} \partial_y \psi
	+ \frac{S}{\Sigma}  \psi \label{eq:D1}
\end{align}
where the superscript $(i)$ indicates the order in $\lambda$. The expressions of $P,Q,S$ can be found in App.~\ref{app:PQS}. We have also suppressed the indicators for $m$ and $\omega$ to prevent notational clutter, which we will continue to do from this point forward. We assume that the solutions for $\psi$ and $\omega$ are close to their GR values, and apply the expansions:
\begin{align}
	\psi(r,y)&=\psi^{(0)}(r,y)+\lambda \psi^{(1)}(r,y)+\mathcal{O}(\lambda^2),
    \label{eq:psi_expanded}\\
    \omega&=\omega^{(0)}+\lambda \omega^{(1)}+\mathcal{O}(\lambda^2).
    \label{eq:om_expanded}
\end{align}
Under this assumption, the EOM for $\psi$ in Eq.~\eqref{eq:D_m_omega} reduces to the perturbative system of equations
\begin{align}
    \mathcal{D}^{(0)}(\omega^{(0)})[\psi^{(0)}] =& 0, 
    \label{eq:Eq_O0}\\
    \mathcal{D}^{(0)}(\omega^{(0)})[\psi^{(1)}] =&-\omega^{(1)}\delta\mathcal{D}^{(0)}[\psi^{(0)}]-\mathcal{D}^{(1)}(\omega^{(0)})[\psi^{(0)}]  \label{eq:Eq_O1},
\end{align}
where we have introduced a new operator $\delta\mathcal{D}^{(0)}=\left.\left(\partial/{\partial\omega}\:\mathcal{D}^{(0)}\right) \right\lvert_{\omega=\omega^{(0)}}$. This system of equations can be solved by first solving Eq.~\eqref{eq:D0} in GR and then using the solutions 
$\omega^{(0)}$ and $\psi^{(0)}$, which enter as source terms in the first-order equation \eqref{eq:Eq_O1}.

At this point, it is important to note that the zeroth-order operator for Kerr in Eq.~\eqref{eq:D0} is separable. We can make the ansatz $\psi^{(0)}(r,y)=R(r)\Theta(y)$, such that the operator splits up in an angular and radial operator:
\begin{equation}
    \mathcal{D}^{(0)}\equiv\frac{1}{\Sigma}(\mathcal{L}_y+\mathcal{L}_r).
\end{equation}
These give rise to an angular and radial equation \cite{teukolskyPerturbationsRotatingBlack1973, leaverAnalyticRepresentationQuasinormal1985}:
\begin{align}
    \mathcal{L}_y[\Theta(y)] &\equiv \frac{d}{d y}(1-y^2)\frac{d}{d y}\Theta(y)\nonumber\\
    &+\Big(A_{lm}+a^2\omega^2y^2 
    -\frac{m}{1-y^2}\Big)\Theta(y)=0, 
    \label{eq:ATE}\\
    \mathcal{L}_r[R(r)] &\equiv \frac{d}{d r} \Delta \frac{d}{d r}R(r)
    +\frac{1}{\Delta}\Big(a^2m^2-4amMr\omega -A_{lm}\nonumber\\
    &+\omega^2(r^4+a^2r^2+2a^2rM)\Big)R(r)=0,
    \label{eq:RTE}
\end{align}
which are coupled by the separation constant $A_{lm}$, which are labelled by a new mode index $l$. At first-order in $\lambda$ the picture gets more complicated, since the operator $\mathcal{D}^{(1)}$ is not separable anymore. 

Now that we have covered the expansion of the EOM, we can address the BCs. A solution to Eqs.~\eqref{eq:Eq_O0}--\eqref{eq:Eq_O1} is not necessarily a QNM: it must be ingoing at the event horizon and outgoing at infinity. We want to enforce this behaviour by explicitly factoring out the right singular behaviour at the boundaries. Hence, we write
\begin{equation}
    \psi(r,y)=A(r,y)\tilde{\psi}(r,y),
\end{equation}
where $A$ is a function that encodes the proper singular behaviour corresponding to the BCs, and $\tilde{\psi}$ is a function that is smooth on the full domain. Note that $A$ also depends on the eigenvalues $m$ and $\omega$. In GR, i.e. for the operator $\mathcal{D}^{(0)}$, the singular structure is well understood: the angular Eq.~\eqref{eq:ATE} has two regular singular points $y=\pm1$ and the radial Eq.~\eqref{eq:RTE} has two regular singular points at the horizons $r=r_\pm$ and an irregular singular point at $r\to\infty$. From a Frobenius analysis it follows that the scalar QNMs should satisfy \cite{leaverAnalyticRepresentationQuasinormal1985, borissovExactSolutionsTeukolsky2010, cookGravitationalPerturbationsKerr2014}:
\begin{numcases}{\psi(r,y)\sim}
    (r-r_+)^{-i\sigma_+}, & for $r\to r_+$,\\
    r^{-1+2iM\omega}e^{\zeta r}, & for $r\to \infty$,
\end{numcases}
where we defined the exponents
\begin{equation}
    \sigma_+=\frac{\omega-m\Omega_H^{(0)}}{2\kappa^{(0)}}, \quad \zeta = i\omega.
\end{equation}
The exponent $\sigma_+$ is a function of the horizon angular velocity and surface gravity for Kerr, which are respectively given by
\begin{equation}
    \Omega_H^{(0)}=\frac{a}{2Mr_+}, \quad \kappa^{(0)}=\frac{r_+-r_-}{4Mr_+}.
\end{equation}
The correct behaviour for the radial solution $R(r)$ is captured by the function \cite{leaverAnalyticRepresentationQuasinormal1985}:
\begin{equation}\label{eq:A0_radial}
    A_r(r)\equiv e^{i\omega r}(r-r_+)^{-i\sigma_+}(r-r_-)^{-1 + 2iM\omega +i\sigma_+}.
\end{equation}
To ensure that $\Theta(y)$ is finite at $y=\pm1$, we take 
\begin{equation}
    A_y(y)\equiv(1-y)^{|m|/2}(1+y)^{|m|/2}.
\end{equation}
Together, these two functions define the zeroth-order regulator $A^{(0)}(r,y)\equiv A_y(y)A_r(r)$. This function can be used to define a new operator 
\begin{equation}\label{eq:D0_reg}
    \tilde{\mathcal{D}}^{(0)}[\tilde{\psi}]\equiv\frac{1}{A^{(0)}}\mathcal{D}^{(0)}[A^{(0)}\tilde{\psi}].
\end{equation}
The transformed operator $\tilde{\mathcal{D}}^{(0)}$ has regular coefficients on the domain, and smooth solutions $\tilde{\psi}$ that do not vanish at the boundaries correspond to fields satisfying the QNM boundary conditions. 

Additionally, we also compactify the radial coordinate to facilitate numerical calculations. In line with \cite{miguelEFTCorrectionsScalar2024, leaverAnalyticRepresentationQuasinormal1985}, we will use the following coordinate \footnote{Note this is different from the coordinate $x\equiv \frac{2r_+}{r}-1$ used in \cite{lamAnalyticAccurateApproximate2026, lamSpinningBlackHoles2026, fernandesLeadingEffectiveField2025}.}: 
\begin{equation}
    z \equiv \frac{r-r_+}{r-r_-}.
\end{equation}
The coordinate maps the outer horizon $r_+$ to $z=0$ and infinity $r\to\infty$ to $z=1$.

Now we turn to the first-order equation \eqref{eq:Eq_O1}. The LHS of this equation, being the Kerr operator, is still regular when we apply the regulator $A^{(0)}$. However, the source terms on the RHS will exhibit irregular coefficients when we use the zeroth-order function $A^{(0)}$, due to the horizon and asymptotic structure being perturbed in the corrected geometry. To solve this problem with divergent coefficients, we introduce a linear regulator $A^{(1)}$ and define the following perturbative solutions
\begin{equation}
    \psi^{(0)}\equiv A^{(0)}\tilde{\psi}^{(0)}, \quad \psi^{(1)}\equiv A^{(0)}\tilde{\psi}^{(1)} + A^{(1)}\tilde{\psi}^{(0)},
\end{equation}
where both $\tilde{\psi}^{(0)}$ and $\tilde{\psi}^{(1)}$ are smooth on the compact domain. This is equivalent to the statement that boundary conditions change when going beyond GR. In this way, we can rewrite the system in Eqs.~\eqref{eq:Eq_O0}--\eqref{eq:Eq_O1} as
\begin{widetext}
\begin{align}
        \tilde{\mathcal{D}}^{(0)}(\omega^{(0)})[\tilde{\psi}^{(0)}] & =0, 
        \label{eq:Eq_O0_reg}\\
        \tilde{\mathcal{D}}^{(0)}(\omega^{(0)})[\tilde{\psi}^{(1)}] & = -\tilde{\mathcal{D}}^{(0)}(\omega^{(0)})[ B^{(1)}\tilde{\psi}^{(0)}] - \omega^{(1)}\delta\tilde{\mathcal{D}}^{(0)}[\tilde{\psi}^{(0)}]-\tilde{\mathcal{D}}^{(1)}(\omega^{(0)})[\tilde{\psi}^{(0)}],
        \label{eq:Eq_O1_reg}
\end{align}
\end{widetext}
where we have ``regularized'' all operators with respect to $A^{(0)}$, similar to Eq.~\eqref{eq:D0_reg}. Furthermore, we have introduced the ratio 
\begin{equation}
    B^{(1)}(z,y)\equiv \frac{A^{(1)}(z,y)}{A^{(0)}(z,y)}.
\end{equation}
The term involving this ratio is ultimately what allows us to properly regularize the first-order equation. The problem is that the source terms introduce poles of the form $(z-1)^{-1}$ and $(z-1)^{-2}$, corresponding to spatial infinity, and poles $z^{-1}$, corresponding to the horizon.  By choosing a proper regulator $A^{(1)}$, we can introduce new poles $(z-1)^{-1}$, $(z-1)^{-2}$, and $z^{-1}$ that will exactly cancel the divergent coefficients, introduced by the other source terms. This way, we ensure that Eq.~\eqref{eq:Eq_O1_reg} is still regular over the full domain $(z,y)\in[0,1]\times[-1,1]$. 

To find the correct regulator $A^{(1)}(z,y)$, we first note that we do not observe any divergences in the angular sector in Eq~\eqref{eq:Eq_O1_reg}. Therefore, we only have to adjust the radial structure of $A^{(1)}$. In this sector, the boundary conditions for Kerr are dependent on the horizon angular velocity and surface gravity, which are modified \cite{canoLeadingHigherderivativeCorrections2019, canoParametrizedQuasinormalMode2024, canoHigherderivativeCorrectionsKerr2024, chungQuasinormalModeFrequencies2024}. These corrections were computed in the context of parametric deviations in the Teukolsky equation in  \cite{canoParametrizedQuasinormalMode2024}. Although we do not directly rely on these analytical solutions, we do find that a similar structure can be used. We also note that we will have to modify the exponent $\zeta$ to account for the modified behaviour at infinity. Similar to findings in \cite{canoLeadingHigherderivativeCorrections2019, chungQuasinormalModeFrequencies2024}, we conclude that the regulator $A^{(1)}$ takes a similar form to $A^{(0)}$ and we only need to adjust the exponents. The ratio will hence be of the form 
\begin{equation}
    B^{(1)}\sim(r-r_+)^{\lambda a_1}(r-r_-)^{\lambda a_2}e^{\lambda a_3r},
\end{equation}
for some unknown exponents $a_{1,2,3}$. Linearizing this in $\lambda$ and switching to the coordinate $z$, this results in the following ansatz for the ratio 
\begin{equation}
    B^{(1)}=C_1 \ln(z) + C_2 \ln(z-1) + C_3 \frac{zr_-  - r_+}{z-1},
\end{equation}
where $C_1$, $C_2$ and $C_3$ are constants, which could still depend on $y$. Acting with $\tilde{\mathcal{D}}^{(0)}$ on the field $B^{(1)}\tilde{\psi}^{(0)}$, the logarithms cancel, and due to derivatives we obtain poles of the form $(z-1)^{-1}$, $(z-1)^{-2}$, $z^{-1}$ and $z^{-2}$ in the coefficients. Enforcing that these cancel the poles in the $\delta\tilde{\mathcal{D}}^{(0)}$ and $\tilde{\mathcal{D}}^{(1)}$ terms, we obtain a system of equations for $C_1$, $C_2$ and $C_3$ that can be solved analytically. This way, we ensure that Eq.~\eqref{eq:Eq_O1_reg} is regular. Determining the structure of poles and cancelling them with a regularizing factor corresponds to enforcing modified boundary conditions.

To summarize, from the Klein-Gordon equation for scalar perturbations we have obtained a regular system of PDEs in Eqs.~\eqref{eq:Eq_O0_reg}--\eqref{eq:Eq_O1_reg} that satisfies the correct BCs. Any smooth solution $\tilde{\psi}$ to these equations is guaranteed to be a scalar QNM of the modified Kerr BH given in Eq.~\eqref{eq:background}. This form is well-suited to be solved numerically using pseudo-spectral techniques.

\section{Numerical pseudo-spectral methods}\label{sec:spectral}
The regularized perturbative equations for scalar QNMs, Eqs.~\eqref{eq:Eq_O0_reg}–\eqref{eq:Eq_O1_reg}, define an eigenvalue problem for $(\omega^{(0)}, \omega^{(1)})$ with corresponding eigenvectors $(\tilde{\psi}^{(0)},\tilde{\psi}^{(1)})$ on the compact domain $(z,y)\in[0,1]\times[-1,1]$. Since all singular behaviour has been factored out analytically, the resulting operators have smooth coefficients, making the system well-suited for pseudo-spectral collocation methods \cite{grandclementSpectralMethodsNumerical2009,Fernandes:2022gde, diasNumericalMethodsFinding2016, canutoSpectralMethodsFundamentals2006}. Our numerical implementation closely follows the pseudo-spectral framework developed in \cite{miguelEFTCorrectionsScalar2024}, but with adapted first-order operators and different normalization conditions to remove gauge freedoms.

We discretize the problem using a Chebyshev pseudo-spectral collocation scheme. The solution is approximated by a bivariate polynomial in $z$ and $y$ and represented in the cardinal basis associated with Chebyshev–Gauss–Lobatto collocation points (see Sec. 2.4 of \cite{canutoSpectralMethodsFundamentals2006}). Derivatives are evaluated through the corresponding Chebyshev differentiation matrices, so that the continuous PDE system is mapped to a generalized matrix eigenvalue problem.

In the radial direction, we employ the mapped Chebyshev–Gauss–Lobatto nodes
\begin{equation}
    z_i=\frac{1}{2}[1-\cos(i\pi/N_r)], \quad i=0,\ldots,N_r
\end{equation}
which cluster near the boundaries $z=0$ and $z=1$. In the angular direction we use
\begin{equation}
    y_j = \cos(j\pi/N_y), \quad j=0,\ldots,N_y.
\end{equation}
corresponding to the standard Chebyshev–Gauss–Lobatto grid on $[-1,1]$. This discretization yields exponential convergence for sufficiently smooth solutions, as expected for pseudo-spectral methods \cite{boydChebyshevFourierSpectral2001}. 

\subsection{The zeroth-order equation}
As exhibited in Eqs.~\eqref{eq:ATE}--\eqref{eq:RTE}, the zeroth-order equation Eq.~\eqref{eq:Eq_O0_reg} separates in a radial and angular equation. This allows us to simplify the collocation method and apply it to the radial and angular sector separately. We introduce the vector $\mathbf{R}$ of length $N_r$ defined by $\mathbf{R}_i\equiv R(z_i)$, and $\mathbf{\Theta}$ of length $N_y$ given by $\mathbf{\Theta}_j\equiv\Theta(y_j)$. By replacing derivatives by their corresponding differentiation matrix (see Sec. 6 of \cite{trefethenSpectralMethodsMATLAB2000} for concrete expressions) and evaluating the coefficients on their respective grids, we discretize the operators $\tilde{\mathcal{L}}_r$ and $\tilde{\mathcal{L}}_y$. We obtain matrices $\mathbf{L}_r$ with shape $N_r\times N_r$ and $\mathbf{L}_y$ with shape $N_y\times N_y$ that are parametrized by $m$ and $a$ and depend smoothly on the eigenvalues $\omega^{(0)}$ and $A_{lm}$. Hence, we can approximate the zeroth-order EOM by 
\begin{numcases}{}
    \mathbf{L}_r(\omega^{(0)},A_{lm})\cdot\mathbf{R} & $=0$,
    \label{eq:RTE_discrete}\\
    \mathbf{L}_y(\omega^{(0)},A_{lm})\cdot\mathbf{\Theta} & $=0$,
    \label{eq:ATE_discrete}
\end{numcases}
The equations are still coupled by their shared eigenvalues $(\omega^{(0)},A_{lm})$, which prevents the system from readily being solved. However, it can be solved by first considering the Schwarzschild limit \cite{miguelEFTCorrectionsScalar2024, mamaniRevisitingQuasinormalModes2022}: for $a\to 0$, the angular equation is solved analytically by spherical harmonics and the separation reduces to the analytical expression $A_{lm}=l(l+1)$. In this case, Eq.~\eqref{eq:RTE_discrete} can be solved using standard numeric eigenvalue solvers. 

To solve Eqs.~\eqref{eq:RTE_discrete}--\eqref{eq:ATE_discrete} for arbitrary spin $a$, we consecutively find new roots for increased spin by using a Newton-Raphson method \cite{diasNumericalMethodsFinding2016}. Starting from $a=0$ and updating the spin by $a\to a+\delta a$, we find a root for the system with given $a$ by iteratively solving the relation
\begin{equation}
    \mathbf{J_F}(\mathbf{x}_n)\cdot\delta\mathbf{x}=-\mathbf{F}(\mathbf{x}_n).
\end{equation}
In our set-up, this equation takes the form
\begin{align}
    &\left(
        \begin{matrix}
           \mathbf{L}_y & \mathbf{0} & \partial_\omega \mathbf{L}_y \cdot \mathbf{\Theta} & \partial_A \mathbf{L}_y \cdot \mathbf{\Theta} \\
           \mathbf{0} & \mathbf{L}_r & \partial_\omega \mathbf{L}_r \cdot \mathbf{R} & \partial_A \mathbf{L}_r \cdot \mathbf{R} \\
           2\mathbf{\Theta}^T & \mathbf{0} & \mathbf{0} & \mathbf{0}\\
           \mathbf{0} & 2 \mathbf{R}^T & \mathbf{0} & \mathbf{0}
        \end{matrix}
    \right) \cdot
    \left(
        \begin{matrix}
            \delta \mathbf{\Theta} \\
            \delta \mathbf{R} \\
            \delta \omega^{(0)} \\
            \delta A_{lm}
        \end{matrix}
    \right) \nonumber\\
    &= -
    \left(
        \begin{matrix}
            \mathbf{L}_y\cdot\mathbf{\Theta}\\
            \mathbf{L}_r\cdot\mathbf{R} \\
            \mathbf{\Theta}^T\cdot\mathbf{\Theta}-1\\
            \mathbf{R}^T\cdot\mathbf{R}-1
        \end{matrix}
    \right).\label{eq:NewtonRaphson}
\end{align}
We solve the resulting linear system for $\delta\mathbf{x}=(\delta\mathbf{\Theta},\delta\mathbf{R},\delta\omega^{(0)},\delta A_{lm})$, using Mathematica's \texttt{LinearSolve}, and update the solution according to
\begin{equation}
\mathbf{x}_{n+1}=\mathbf{x}_n+\delta\mathbf{x}.
\end{equation}
In our set-up, the iteration is repeated until the residual $|\delta \omega| < 10^{-p}$, consistent with the working precision $p$. 

The off-diagonal terms in the Jacobian in Eq.~\eqref{eq:NewtonRaphson} encode the parametric dependence of the operators on the eigenvalues and eliminate the scaling degeneracy in the eigenvalue problem. The upper right block takes care of the parametric dependence through derivatives with respect to the eigenvalues. These derivatives are calculated by first taking an analytic derivative of the continuous operators and then discretizing the resulting operators.

The lower block enforces a normalization of the eigenvectors and, consequently, removes the scaling degeneracy. We impose
\begin{equation}
    \mathbf{\Theta}^T\mathbf{\Theta}=1, \quad \mathbf{R}^T\mathbf{R}=1.
\end{equation}
This removes null directions from the Jacobian and renders Eq.~\eqref{eq:NewtonRaphson} non-singular in a neighbourhood of the solution.

For the fundamental modes, we start from known values (e.g. \cite{bertiGravitationalwaveSpectroscopyMassive2006}) as initial guesses for $(\omega^{(0)}, A_{lm})$, and a constant array for $\mathbf{R}$ and a spherical harmonic for $\mathbf{\Theta}$. For a given spin, we use the iterative Newton-Raphson method to find correct solutions for $\mathbf{R}$ and $\mathbf{\Theta}$, and $(\omega^{(0)}, A_{lm})$ at the desired precision. The overtones are more sensitive, since the Newton-Raphson root finder might accidentally recover the fundamental mode frequency that lies close. To prevent this, we initialize the eigenvector $\mathbf{R}$ with the eigenvector constructed using a Leaver solver \cite{leaverAnalyticRepresentationQuasinormal1985} for this overtone, instead of a constant array. This allows us to track the overtones correctly, using a similar procedure as the fundamental tones. By incrementing the spin, we can find the zeroth-order solutions which are then used as input for the correction equations.

\subsection{The first-order equation}
At first-order in the perturbative expansion in Eq.~\eqref{eq:Eq_O1_reg}, the separability of the zeroth-order problem is lost and the equation must be solved on the full two-dimensional domain. The discretized operators therefore act on vectors of dimension $N_r \cdot N_y$, and all matrices appearing in this section are of size $(N_r\cdot N_y)\times (N_r \cdot N_y)$ unless stated otherwise.

The zeroth-order solution of the separable problem is constructed from the tensor product of the radial and angular eigenvectors,
\begin{equation}
    \boldsymbol{\tilde{\Psi}}^{(0)} = \mathbf{R} \otimes \mathbf{\Theta},
\end{equation}
which we flatten into a single column vector of length $N_r\cdot N_y$, using Mathematica's \texttt{Flatten} function. Since the first-order equation is not separable, the correction $\boldsymbol{\tilde{\Psi}}^{(1)}$ is treated as a general vector of length $N_r\cdot N_y$ that is to be determined.

Derivatives on the two-dimensional grid are represented using tensor products of the one-dimensional Chebyshev differentiation matrices. Denoting the first-derivative matrices in the radial and angular directions by $\mathbf{D}_z$ and $\mathbf{D}_y$, respectively, we construct
\begin{align}
    \partial_z & \to \mathbf{D}_z \otimes \mathbf{I}_y, \\
    \partial_y & \to \mathbf{I}_r \otimes \mathbf{D}_y,
\end{align}
with $\mathbf{I}_{r,y}$ the identity matrices in the corresponding subspaces. Higher derivatives are obtained analogously. In this way, all differential operators inherit a natural Kronecker-product structure. By using these differential matrices, by expressing the coefficients on the Chebyshev grids, and by multiplying constants like $a$, $m$, $\omega^{(0)}$ with the identity matrix, the operators are discretized. Note that only the regular parts of $\delta\tilde{\mathcal{D}}^{(0)}$ and $\tilde{\mathcal{D}}^{(1)}$ appear. Any divergent coefficients are cancelled exactly by the term involving the ratio $B^{(1)}$. Hence, we only have the regular, discretized operators $\mathbf{\tilde{D}}^{(0)}$, $\boldsymbol{\delta}\mathbf{\tilde{D}}^{(0)}$, and $\mathbf{\tilde{D}}^{(1)}$ appearing. Now we can express the non-homogeneous Eq.~\eqref{eq:Eq_O1_reg} as the linear problem:
\begin{align}
    &\left(
        \begin{matrix}
           \mathbf{\tilde{D}}^{(0)} & (\boldsymbol{\delta}\mathbf{\tilde{D}}^{(0)}\cdot \boldsymbol{\tilde{\Psi}}^{(0)})^T\\
           (\boldsymbol{\tilde{\Psi}}^{(0)})^T & 0
        \end{matrix}
    \right) \cdot
    \left(
        \begin{matrix}
            \boldsymbol{\tilde{\Psi}}^{(1)} \\
            \omega^{(1)}
        \end{matrix}
    \right) \nonumber\\
    &=
    \left(
        \begin{matrix}
            -\mathbf{\tilde{D}}^{(1)}\cdot\boldsymbol{\tilde{\Psi}}^{(0)}\\
            0
        \end{matrix}
    \right),
\end{align}
which we will solve for $(\boldsymbol{\tilde{\Psi}}^{(1)}, \omega^{(1)})^T$. We have introduced a new row to remove the freedom 
$\boldsymbol{\tilde{\Psi}}^{(1)} \rightarrow \boldsymbol{\tilde{\Psi}}^{(1)} + \alpha \boldsymbol{\tilde{\Psi}}^{(0)}$ and enforce solvability. The additional row enforces the orthogonality constraint
\begin{equation}
    (\boldsymbol{\tilde{\Psi}}^{(0)})^T \boldsymbol{\tilde{\Psi}}^{(1)} = 0.
\end{equation}
This condition fixes the first-order correction uniquely within the complement of the zeroth-order eigenspace, and consequently, renders the augmented matrix nonsingular. The system can therefore be solved using Mathematica's \texttt{LinearSolve} to obtain both the first-order eigenfunction correction and the frequency shift $\omega^{(1)}$.

\section{Results}\label{sec:results}

We have calculated results for scalar QNM corrections in an EFT of gravity for modes $l\le5$ and all corresponding $m$ for spins up to $a/M=0.99$, demonstrating the viability of pseudo-spectral methods for QNM calculations in higher derivative theories. The corrections to the scalar QNM frequencies obtained in this study are available in a public GitHub repository \footnote{\faGithub\hspace{0.2em}\href{https://github.com/StefHusken/scalar-QNMs-higher-derivative-gravity}{scalar QNMs higher derivative gravity}}. In what follows, we set $M=1$ without loss of generality.

We present our results using both the numerically computed corrected Kerr metric from Ref.~\cite{fernandesLeadingEffectiveField2025} and the perturbative spin expansion of Ref.~\cite{canoLeadingHigherderivativeCorrections2019} to make a comparison of their accuracy across different spins. The numerical background is given with spectral order $N_r=50$, $N_y=16$ for spins up to $a=0.6$. Higher spins use a numerical resolution of $N_r=60$, $N_y=24$. Only limited resolution is needed in the angular sector. We compare to a spin expansion up to $a^{14}$, which remains accurate up until $a\approx0.75$ \cite{canoLeadingHigherderivativeCorrections2019}.

\begin{figure}
\centering
\includegraphics[width=\linewidth]{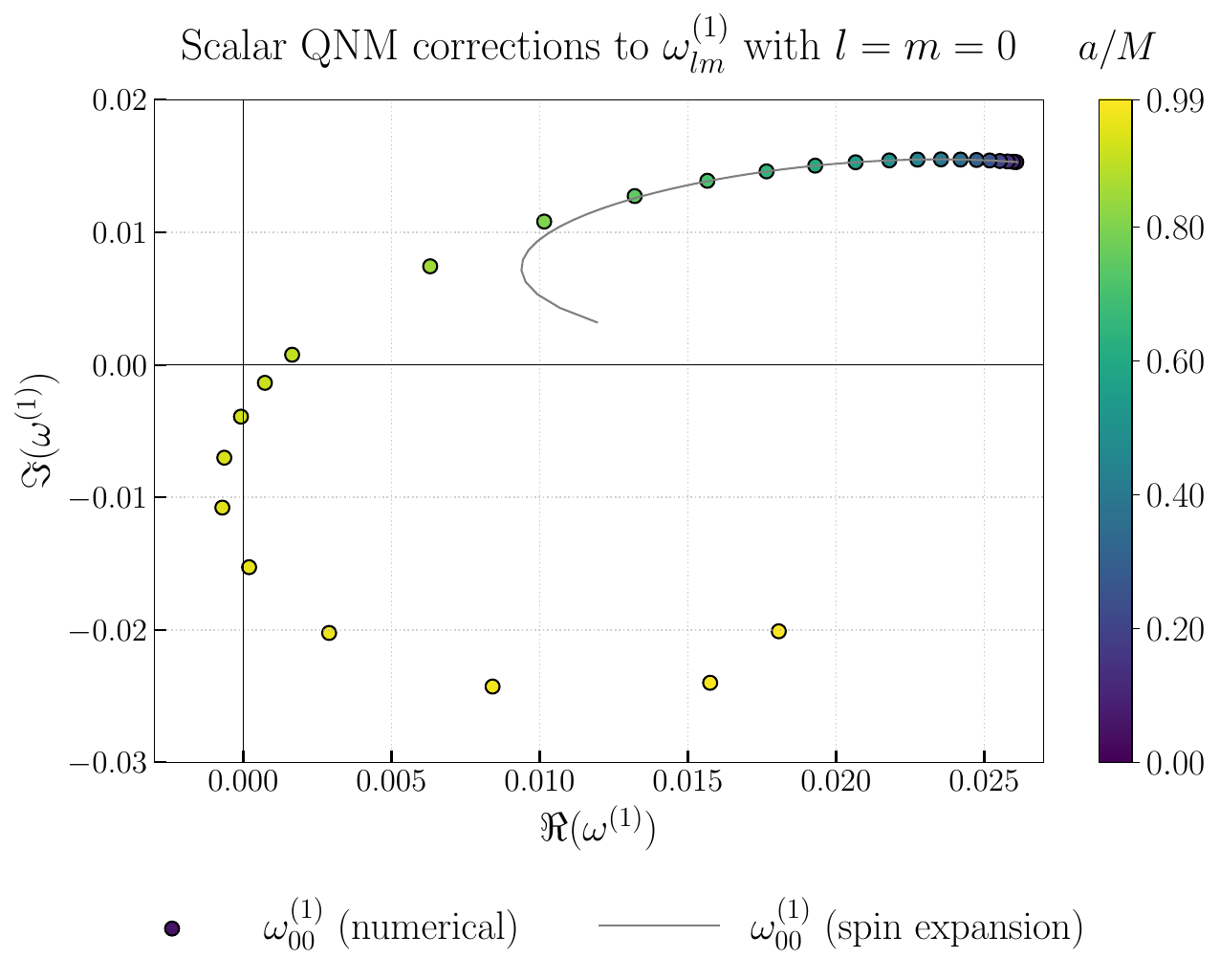}
\caption{Lowest order corrections of scalar QNMs for $l=m=0,$ shown in the complex plane for spins up to $a=0.99$. Results from the numerical background (circle markers) are shown with increments $\Delta a=0.05$ up to $a=0.9,$ and $\Delta a=0.01$ thereafter. The solid line denotes the results from the spin expansion, up until $a=0.9.$
}
\label{fig:l0s0cubic}
\end{figure}

Figure~\ref{fig:l0s0cubic} depicts the corrections to the scalar QNM frequencies for the $l=m=0$ mode in the complex plane for increasing spins. The figure shows the resulting corrections for both the spin expansion as well as the numerical background. As can be seen, both backgrounds yield comparable results for moderate spin but for large spins the spin expansion fails to be accurate, as expected. The corrections from the numerical background are considerably bigger for large spins. A collocation grid size of $N_r=N_y=120$ has been used to solve for $\omega^{(1)}$, leading to relative errors for $\omega^{(1)}$ for all spins smaller than $10^{-4}$.

\begin{figure*}
\centering
\includegraphics[width=\linewidth]{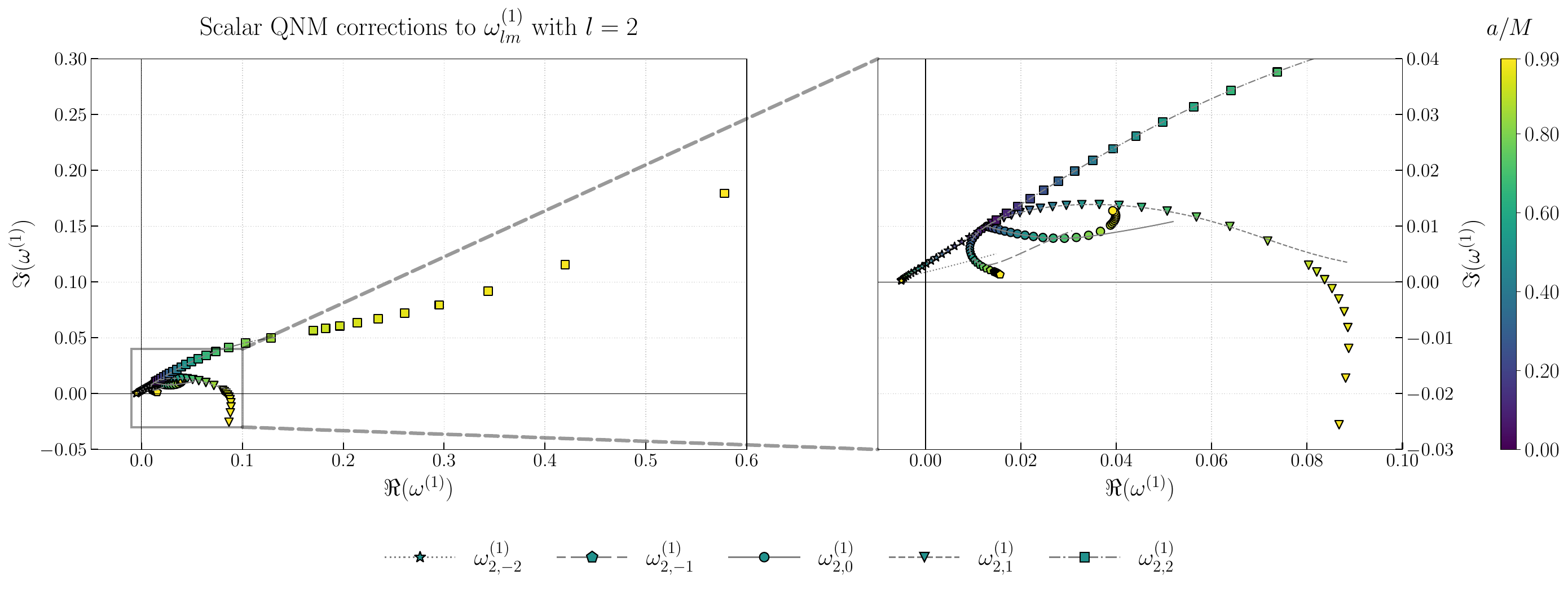}
\caption{Lowest order corrections of scalar QNMs in cubic gravity for $l=2,$ shown in the complex plane, for spins up to $a=0.99$. Results from the numerical background, depicted by the markers, are shown with increments of $\Delta a=0.05$ up to $a=0.9,$ and $\Delta a=0.01$ thereafter. The $m=2$ mode is truncated in the plot, as the corrections get an order of magnitude bigger for larger spins. The solid gray lines denotes the results based on the spin expansion, up until $a=0.9$.
}
\label{fig:l2s0cubic}
\end{figure*}

Fig.~\ref{fig:l2s0cubic} presents the $l=2$ modes. We have adopted the same grid with size $N_r=N_y=120$, which leads to relative errors for all spins smaller than $10^{-8}$ for these modes. Again, we compare the results between the numerical and analytical background, which exhibits excellent agreement for low and moderate spins. At spins $a\gtrsim0.8$, we start to observe significant differences as the spin expansion becomes inaccurate. Most importantly, the spin expansions fail to capture the right trend in many cases and develop in the wrong direction in the complex plane.

For the overtones, we observe similar behaviour. The spin expansion is able to accurately capture their behaviour up to $a\sim0.8$, but for higher spins it fails to produce the right evolution in spin. Furthermore, we find that the corrections to the imaginary part of the overtone frequencies, $\Im(\omega^{(1)}_{lm1})$, are systematically larger than those of the corresponding fundamental modes. For several modes, the overtone corrections exceed the fundamental-mode corrections by approximately an order of magnitude. This behaviour is consistent with previous studies indicating that overtone modes are more sensitive to short-range modifications of the near-horizon geometry than the fundamental mode \cite{silvaQuasinormalModesExcitation2024, konoplyaFirstFewOvertones2024}. The enhanced modification of the overtones therefore suggests that they provide a particularly sensitive probe of the higher-curvature corrections encoded in the EFT.

As the spin increases above $a>0.9$, certain modes receive corrections that are orders of magnitude larger than the corrections in the slowly spinning regime, notably the $m=2$ mode for $l=2$ in Fig.~\ref{fig:l2s0cubic}. We find a similar trend for $l=1$ at $m=1$, $l=3$ at $m=3$, for $l=4$ at $m=3,4$, and for $l=5$ at $m=3,4,5$. The corresponding overtones show the same type of behaviour. This amplification of beyond-GR effects in the large-spin regime highlights the observational importance of rapidly spinning BHs.

\begin{figure}
\centering
\includegraphics[width=\linewidth]{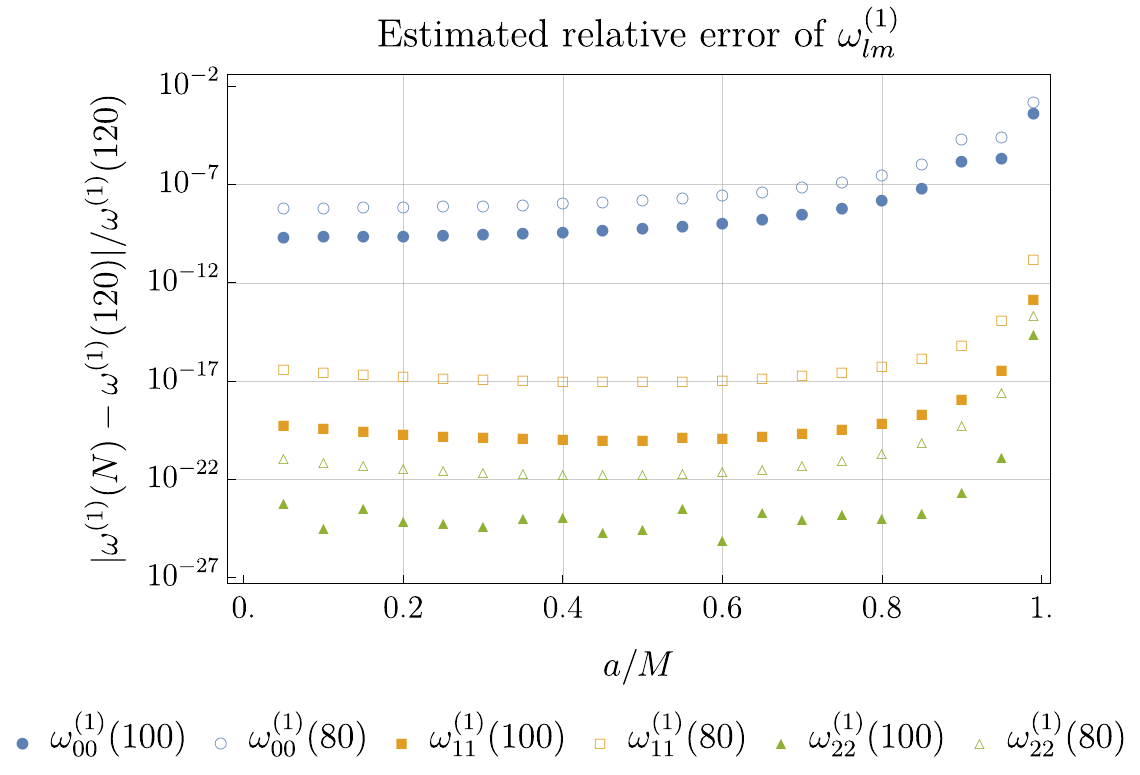}
\caption{Estimated relative errors for corrections $\omega^{(1)}$ for $l=m$ modes with $0\le l \le 2$ as a function of spin. Filled markers indicate the relative errors between high resolution $N_r=N_y=120$ and low  resolution $N_r=N_y=80$, and filled markers between the same high resolution and moderate resolution $N_r=N_y=100$.
}
\label{fig:cubicconvergence}
\end{figure}

We have verified the accuracy of our numerical methods in several ways. First, we have computed the corrections for a given spin at different resolutions to check their convergence. Relative errors are estimated by comparing results at a given grid size to those obtained with a reference grid of $N_r=N_y=120$. As shown by the relative errors for a selection of modes in Fig.~\ref{fig:cubicconvergence}, increasing the resolution of the grid decreases the error in the results. Errors become at least an order of magnitude smaller. This demonstrates that the results are numerically stable. In particular, it also shows that the amplification of e.g. the $l=m=2$ mode is not a numerical artifact. Furthermore, we have compared results at lows spins to existing methods. Our results agree with values in \cite{canoRingingRotatingBlack2020} up to the specified uncertainty and we have also verified the results by using an eigenvalue perturbation method based on the spin expansion, which shows excellent agreement \cite{zimmermanQuasinormalModesKerr2015, hussainApproachComputingSpectral2022,canoQuasinormalModesRotating2023}.

\section{Conclusion and Discussion}\label{sec:discussion}

In this work, we have computed the corrections to the scalar QNM spectrum in cubic higher-derivative gravity. We have been able to accurately capture the behaviour of all fundamental modes $l\le5$ and $|m|\le l$ and the first overtone of $2 \le l \le 5$ for spins up to $a/M\le0.99$. The methods provided in this paper are general for any higher-derivative gravity (see also \cite{huskenQuadraticGravityCorrections2026}), as long as the higher derivative effects are treated perturbatively as a correction to Kerr. These scalar QNMs can be used as a proxy for the gravitational QNMs \cite{paniScalarShortcutBeyondKerr2026}. They open the door to a systematic study of beyond-GR corrections to QNMs across a broad class of theories, and to placing observational constraints on higher-derivative gravity using GW observations of rapidly spinning black holes.

Our results confirm that the spin expanded background becomes inaccurate at large spins and does not correctly predict the behaviour of corrections to the QNM frequencies for $a\gtrsim0.8$. This demonstrates the necessity of using numerical backgrounds for studying rapidly spinning BHs beyond GR. Interestingly, limited resolution suffices for the numerical background, with 24 basis functions being the highest spectral order. This is indicative of the better-conditioned nature of numerical solutions compared to spin expansions, that deteriorate quickly at large spins. The accuracy of the frequency corrections is primarily limited by the grid size of the pseudo-spectral solver.

We observe, in general, that corrections become larger as spin increases, which underlines the importance of understanding the QNM modifications for highly spinning BHs. Certain modes grow by orders of magnitude as we approach the near-extremal regime. Studies of the corrections in the Eikonal limit for quartic higher-derivative gravity suggest that the growth saturates and corrections decrease in magnitude again closer to extremality, rendering the corrections finite \cite{canoEikonalQuasinormalModes2025}. Future work should push our methods to higher spins to examine the near-extremal regime in greater detail and test this conjecture.

The large amplification of some modes also raises questions about the validity of the lowest order approximation in Eqs.~\eqref{eq:psi_expanded} and \eqref{eq:om_expanded}, and the regime of validity of the EFT. In \cite{canoAmplificationNewPhysics2025} it is suggested that these amplifications can occur within the validity of the EFT. This can be understood from the non-perturbative nature of a shift of the phase boundary between damped modes and zero-damped modes in the extremal limit \cite{yangBranchingQuasinormalModes2013, yangQuasinormalModesNearly2013}. Near this boundary, the lowest order approximation might fail and one has to carefully perform an simultaneous expansion in both the EFT scale as well as the distance to the phase boundary. This shift of the phase boundary in beyond GR is expected to be a theory independent phenomenon in any higher derivative gravity \cite{canoAmplificationNewPhysics2025}. Similar behaviour is observed in scalar Gauss-Bonnet and dynamical Chern-Simons close to this phase boundary \cite{huskenQuadraticGravityCorrections2026}. 

Both a more thorough treatment of the near-extremal regime and a broader exploration of higher-derivative theories are left for future investigation. Furthermore, the pseudo-spectral framework developed here does not rely on features specific to the scalar wave equation. It appears well-suited for extensions to gravitational perturbations. In the tensorial sector, the metric perturbations couple directly to the modified field equations and parity-violating operators will have to be treated consistently. Addressing these complications and extending the analysis to gravitational QNMs is left for future work.

\section*{Acknowledgments}

We thank Llibert Aresté Saló, Pablo Cano, Adrian Chung, Marina David, Nicola Franchini, Kelvin Lam, Henri Inchausp\'e and Nicol\'as Yunes for insightful discussions and comments on the draft of this manuscript.
T.v.d.S. acknowledges funding from the Research Foundation - Flanders (FWO, Fonds Wetenschappelijk Onderzoek) through a PhD Fellowship under FWO Grant No. 1125026N.
T.v.d.S. and T.H. thank the Belgian Federal Science Policy Office (BELSPO) for the provision of financial support in the framework of the PRODEX Programme of the European Space Agency (ESA) under contract number PEA4000144253.
This research was supported in part by KU Leuven grant C16/25/010.
S.M. and V.C. acknowledge funding from the Simons Foundation International and the Simons Foundation through Grant No. SFI-MPS-BH-00012593-11.
The Center of Gravity is a Center of Excellence funded by the Danish National Research Foundation under grant No. DNRF184.
We acknowledge support by VILLUM Foundation (grant no. VIL37766).
V.C.\ is a Villum Investigator.  
P.~F. acknowledges funding by the Deutsche Forschungsgemeinschaft (DFG, German Research Foundation) under Germany’s Excellence Strategy EXC 2181/1 - 390900948 (the Heidelberg STRUCTURES Excellence Cluster).
T.G.F.L.~is supported by grants from the the Research Foundation – Flanders (FWO, Fonds Wetenschappelijk Onderzoek) Grant No. I002123N and I000725N.
V.C. acknowledges financial support provided under the European Union’s H2020 ERC Advanced Grant “Black holes: gravitational engines of discovery” grant agreement no. Gravitas–101052587. 
Views and opinions expressed are however those of the author only and do not necessarily reflect those of the European Union or the European Research Council. Neither the European Union nor the granting authority can be held responsible for them.
This project has received funding from the European Union's Horizon 2020 research and innovation programme under the Marie Sklodowska-Curie grant agreement No 101007855 and No 101131233.
The Tycho supercomputer hosted at the SCIENCE HPC center at the University of Copenhagen was used for supporting this work.

\bibliography{References}

{\onecolumngrid

\appendix

\clearpage

\section{First-order contributions to Klein-Gordon operator}\label{app:PQS}
In this section, we provide explicit expressions for the corrections to the Klein-Gordon operator on a rotating BH background in higher-derivative gravity. We have verified and corrected minor typographical errors in the expressions with respect to \cite{canoRingingRotatingBlack2020}.
As given in Eq.~\eqref{eq:D1}, the modifications to the BH geometry lead to modifications of the form: 
\begin{equation}
    \mathcal{D}^{(1)} \psi = 
	-\frac{H_3\Delta}{\Sigma}\partial_r^2 \psi 
	+\frac{H_3(y^2-1)}{\Sigma}\partial_y^2 \psi 
	+ \frac{P}{\Sigma} \partial_r \psi 
	+ \frac{Q}{\Sigma} \partial_y \psi
	+ \frac{S}{\Sigma}  \psi
\end{equation}
The functions $P$, $Q$, and $S$ that depend on the metric corrections $H_i(r,y)$ are given by the following expressions \cite{canoRingingRotatingBlack2020}:
\begin{align}
P&=-\frac{H_2^{(1,0)}}{\Sigma^2}4a^2M^2r^2(y^2-1)+2H_3(M-r)+\frac{(\Sigma-2Mr) H_4^{(1,0)}-\Sigma H_1^{(1,0)}}{2\Sigma^2}(a^4y^2+a^2r(-2My^2+2M+r+ry^2)+r^4)\nonumber\\
&+\frac{4a^2(H_4-2H_2)}{\Delta \Sigma^3}M^2r(y^2-1)(a^4y^2-a^2r(My^2+r)+r^3(3M-2r))\nonumber\\
&+\frac{H_1 M}{\Delta\Sigma^2}(-a^6y^2+a^4r^2(1-2y^2)+a^2r^3(4M(y^2-1)-r(y^2-2))+r^6)
\\[1\baselineskip]
Q&=\frac{H_2^{(0,1)}}{\Delta\Sigma^2}4a^2M^2r^2(y^2-1)^2-\frac{2a^2H_1}{\Delta\Sigma^2}Mry(y^2-1)(a^2+r^2)+2H_3y\nonumber\\
&-\frac{(y^2-1)H_4^{(0,1)}}{2\Delta\Sigma^2}(a^6y^4+a^4ry^2(r(y^2+2)-2My^2)+a^2r^2(4M^2(y^2-1)-4Mry^2+r^2(2y^2+1))+r^5(r-2M))\nonumber\\
&+\frac{4a^2(H_4-2H_2)}{\Delta\Sigma^3}M^2r^2y(y^2-1)(a^2(y^2-2)-r^2)+\frac{(y^2-1)H_1^{(0,1)}}{2\Delta\Sigma}(a^4y^2+a^2r(-2My^2+2M+r+ry^2)+r^4)
\\[1\baselineskip]
S&=\frac{4aH_2}{\Delta^2\Sigma^2}Mr\big[2am^2Mr(a^2y^2+r(r-2M))-8a^2mM^2r^2(y^2-1)\omega-m\omega \Delta\Sigma^2\nonumber\\
&+2aMr(y^2-1)\omega^2(a^4y^2+a^2r(-2My^2+2M+r+ry^2)+r^4)\big]\nonumber\\
&+\frac{H_1}{\Delta^2\Sigma}(a^4y^2\omega+a^2r\omega(-2My^2+2M+r+ry^2)-2amMr+r^4\omega)^2\nonumber\\
&-\frac{H_4}{(y^2-1)\Delta^2\Sigma^2}(a^4y^2+a^2r(-2My^2+2M+r+ry^2)+r^4)(a^2my^2-2aMr(y^2-1)\omega+mr(r-2M))^2,
\end{align}
where the superscripts $(i,j)$ indicate the order of the derivative with respect to $r$ and $y$, respectively. Since we solve the Klein-Gordon equation to leading order in $\lambda$, second-order derivatives of $\psi$ can be eliminated using the zeroth-order equation $\mathcal{D}^{(0)}\psi=0$. Furthermore, we can reduce first-order derivatives of the scalar field, by first-order redefinition of the field. This transformation is given by
\begin{equation}
    \psi\to(1+\lambda F(r,y))\psi,
\end{equation}
where the function $F(r,y)$ takes the form
\begin{align}
    F(r,y) &= \frac{8 a^2 M^2 r^2 \left(y^2-1\right)H_2}{4\Delta\Sigma^2}
    +\frac{\left(a^2 y^2+r^2\right)H_1}{4\Delta\Sigma^2}\left(a^4 y^2+a^2 r \left(-2 M y^2+2 M+r y^2+r\right)+r^4\right)\nonumber\\
    &-\frac{H_4}{4\Delta\Sigma^2}\left(a^6 y^4+a^4 r y^2 \left(r \left(y^2+2\right)-2 M y^2\right)+a^2 r^2 \left(4 M^2 \left(y^2-1\right)-4 M r y^2+r^2 \left(2 y^2+1\right)\right)+r^5 (r-2 M)\right).
\end{align}
After the reduction of the derivatives, the first-order operator is consistent with a source term:
\begin{equation}
    \mathcal{D}^{(1)}\psi=\frac{S'}{\Sigma}\psi
\end{equation}
The explicit form of the source term $S'$ is lengthy and is therefore omitted for compactness.

\clearpage

\section{Some tabulated scalar QNM frequencies}
We provide a selection of the scalar QNM frequencies and their cubic higher-curvature corrections in Table~\ref{tab:QNMs}. For the full list of frequencies we refer to the  \href{https://github.com/StefHusken/scalar-QNMs-higher-derivative-gravity}{GitHub Repository \hspace{0.1em}\faGithub}.

\begin{table*}[ht]
\begin{tabular}{ccc|r||r|r|r|r|r|r|r}
\multicolumn{4}{c||}{Parameters}&\multicolumn{7}{c}{$a/M$}\\
\hline\hline
$l$&$m$&$n$&Component&$0.05$&$0.25$&$0.5$&$0.75$&$0.9$&$0.95$&$0.99$\\
\hline\hline
\multirow{4}{*}{$0$}&\multirow{4}{*}{$0$}&\multirow{4}{*}{$0$}
&$\Re(M\omega^{(0)})$ 
&$ 0.110474$ &$ 0.110943 $ &$ 0.112380 $ &$ 0.114322 $ &$ 0.113848 $ &$ 0.111992 $ &$ 0.110447 $\\
&&&$\Im(M\omega^{(0)})$ 
&$ -0.104872$ &$ -0.104289 $ &$ -0.102183 $ &$ -0.097298 $ &$ -0.091569 $ &$ -0.089559 $ &$ -0.089499 $\\
&&&$\Re(M\omega^{(1)})$ 
&$ 0.026079 $ &$ 0.025179 $ &$ 0.021796 $ &$ 0.013206 $ &$ 0.001648 $ &$ 0.000196 $ &$ 0.018066 $ \\
&&&$\Im(M\omega^{(1)})$
&$ 0.015302 $ &$ 0.015419 $ &$ 0.015423 $ &$ 0.012731 $ &$ 0.00075 $  &$ -0.015272 $ &$ -0.020111 $\\
\hline
\multirow{4}{*}{$1$}&\multirow{4}{*}{$1$}&\multirow{4}{*}{$0$}
&$\Re(M\omega^{(0)})$ 
&$ 0.296889 $ &$ 0.314934 $ &$ 0.344753 $ &$ 0.390378 $ &$ 0.437234 $ &$ 0.462261 $ &$ 0.493423 $\\
&&&$\Im(M\omega^{(0)})$ 
&$ -0.097624 $ &$ -0.097005 $ &$ -0.094395 $ &$ -0.086394 $ &$ -0.071848 $ &$ -0.060091 $ &$ -0.0367120 $\\
&&&$\Re(M\omega^{(1)})$ 
&$ 0.011988 $ &$ 0.018980 $ &$ 0.032173 $ &$ 0.058215 $ &$ 0.099776 $ &$ 0.134764 $ &$ 0.205859 $\\
&&&$\Im(M\omega^{(1)})$
&$ 0.014158 $ &$ 0.017886 $ &$ 0.024102 $ &$ 0.033508 $ &$ 0.045878 $ &$ 0.061257 $ &$ 0.159191 $\\
\hline
\multirow{4}{*}{$1$}&\multirow{4}{*}{$-1$}&\multirow{4}{*}{$0$}
&$\Re(M\omega^{(0)})$ 
&$ 0.289168 $ &$ 0.275671 $ &$ 0.344753 $ &$ 0.261572 $ &$ 0.243370 $ &$ 0.241372 $ &$ 0.239809 $\\
&&&$\Im(M\omega^{(0)})$ 
&$ -0.097659 $ &$ -0.097362 $ &$ -0.094395 $ &$ -0.096505 $ &$ -0.094422 $ &$ -0.094124 $ &$ -0.093882 $\\
&&&$\Re(M\omega^{(1)})$ 
&$ 0.009213 $ &$ 0.004714 $ &$ 0.009878 $ &$ -0.001914 $ &$ -0.002716 $ &$ -0.002852 $ &$ -0.002905 $\\
&&&$\Im(M\omega^{(1)})$
&$ 0.012586 $ &$ 0.009878 $ &$ 0.007104 $ &$ 0.004854 $ &$ 0.003791 $ &$ 0.003501 $ &$ 0.003293 $\\
\hline
\multirow{4}{*}{$2$}&\multirow{4}{*}{$2$}&\multirow{4}{*}{$0$}
&$\Re(M\omega^{(0)})$ 
&$ 0.491359 $ &$ 0.585989 $ &$ 0.344753 $ &$ 0.679503 $ &$ 0.781638 $ &$ 0.840982 $ &$ 0.928028 $\\
&&&$\Im(M\omega^{(0)})$ 
&$ -0.096732 $ &$ -0.093494 $ &$ -0.094395 $ &$ -0.085026 $ &$ -0.069289 $ &$ -0.056470 $ &$ -0.031063 $\\
&&&$\Re(M\omega^{(1)})$ 
&$ 0.014795 $ &$ 0.024739 $ &$ 0.044128 $ &$ 0.086293 $ &$ 0.170208 $ &$ 0.260801 $ &$ 0.577936 $\\
&&&$\Im(M\omega^{(1)})$
&$ 0.011106 $ &$ 0.016455 $ &$ 0.026158 $ &$ 0.041235 $ &$ 0.056667 $ &$ 0.072062 $ &$ 0.179495 $\\
\hline
\multirow{4}{*}{$2$}&\multirow{4}{*}{$-2$}&\multirow{4}{*}{$0$}
&$\Re(M\omega^{(0)})$ 
&$ 0.476299 $ &$ 0.450059 $ &$ 0.422751 $ &$ 0.399883 $ &$ 0.387799 $ &$ 0.384002 $ &$ 0.381041 $\\
&&&$\Im(M\omega^{(0)})$ 
&$ -0.096748 $ &$ -0.096432 $ &$ -0.095615 $ &$ -0.094524 $ &$ -0.093790 $ &$ -0.093535 $ &$ -0.093329 $\\
&&&$\Re(M\omega^{(1)})$ 
&$ 0.010910 $ &$ 0.004710 $ &$ -0.000740 $ &$ -0.004072 $ &$ 0.005028 $ &$ -0.005119 $ &$ -0.005070 $\\
&&&$\Im(M\omega^{(1)})$
&$ 0.009004 $ &$ 0.005674 $ &$ 0.002786 $ &$ 0.001005 $ &$ 0.000407 $ &$ 0.000291 $ &$ 0.000234 $\\
\hline
\multirow{4}{*}{$2$}&\multirow{4}{*}{$2$}&\multirow{4}{*}{$1$}
&$\Re(M\omega^{(0)})$ 
&$ 0.472172 $ &$ 0.510210 $ &$ 0.573442 $ &$ 0.671954 $ &$ 0.777683 $ &$ 0.838305 $ &$ 0.926686 $\\
&&&$\Im(M\omega^{(0)})$ 
&$ -0.295312 $ &$ -0.292632 $ &$ -0.283336 $ &$ -0.256350 $ &$ -0.208008 $ &$ -0.169108 $ &$ -0.092600 $\\
&&&$\Re(M\omega^{(1)})$ 
&$ 0.036281 $ &$ 0.052967 $ &$ 0.080069 $ &$ 0.121710 $ &$ 0.184152 $ &$ 0.254202 $ &$ 0.549706 $\\
&&&$\Im(M\omega^{(1)})$
&$ 0.031777 $ &$ 0.046294 $ &$ 0.073297 $ &$ 0.117826 $ &$ 0.168245 $ &$ 0.220386 $ &$ 0.556800 $\\
\hline
\multirow{4}{*}{$3$}&\multirow{4}{*}{$3$}&\multirow{4}{*}{$0$}
&$\Re(M\omega^{(0)})$ 
&$ 0.686858 $ &$ 0.739705 $ &$ 0.828800 $ &$ 0.971226 $ &$ 1.130108 $ &$ 1.224702 $ &$ 1.368634 $\\
&&&$\Im(M\omega^{(0)})$ 
&$ -0.096477 $ &$ -0.095903 $ &$ -0.093235 $ &$ -0.084635 $ &$ -0.068627 $ &$ -0.055646 $ &$ -0.030244 $\\
&&&$\Re(M\omega^{(1)})$ 
&$ 0.020370 $ &$ 0.033406 $ &$ 0.057817 $ &$ 0.109793 $ &$ 0.222300 $ &$ 0.358550 $ &$ 0.907336 $\\
&&&$\Im(M\omega^{(1)})$
&$ 0.009677 $ &$ 0.015469 $ &$ 0.026721 $ &$ 0.045499 $ &$ 0.064244 $ &$ 0.080182 $ &$ 0.187185 $\\
\hline
\multirow{4}{*}{$3$}&\multirow{4}{*}{$-3$}&\multirow{4}{*}{$0$}
&$\Re(M\omega^{(0)})$ 
&$ 0.664437 $ &$ 0.625471 $ &$ 0.585048 $ &$ 0.551306 $ &$ 0.533519 $ &$ 0.527937 $ &$ 0.523586 $\\
&&&$\Im(M\omega^{(0)})$ 
&$ -0.096486 $ &$ -0.096160 $ &$ -0.095345 $ &$ -0.094273 $ &$ -0.093559 $ &$ -0.093312 $ &$ -0.093113 $\\
&&&$\Re(M\omega^{(1)})$ 
&$ 0.015150 $ &$ 0.006643 $ &$ -0.001064 $ &$ -0.005915 $ &$ -0.007294 $ &$ -0.007393 $ &$ -0.007271 $\\
&&&$\Im(M\omega^{(1)})$
&$ 0.007507 $ &$ 0.004230 $ &$ 0.001616 $ &$ 0.000196 $ &$ -0.000177 $ &$ -0.000215 $ &$ -0.000205 $\\
\hline
\multirow{4}{*}{$3$}&\multirow{4}{*}{$3$}&\multirow{4}{*}{$1$}
&$\Re(M\omega^{(0)})$ 
&$ 0.672651 $ &$ 0.727616 $ &$ 0.819827 $ &$ 0.966103 $ &$ 1.127783 $ &$ 1.223412 $ &$ 1.368301 $\\
&&&$\Im(M\omega^{(0)})$ 
&$ -0.291787 $ &$ -0.289871 $ &$ -0.281150 $ &$ -0.254542 $ &$ -0.206014 $ &$ -0.166927 $ &$ -0.090691 $\\
&&&$\Re(M\omega^{(1)})$ 
&$ 0.033850 $ &$ 0.052036 $ &$ 0.083480 $ &$ 0.137952 $ &$ 0.235410 $ &$ 0.355907 $ &$ 0.891710 $\\
&&&$\Im(M\omega^{(1)})$
&$ 0.028481 $ &$ 0.045112 $ &$ 0.077457 $ &$ 0.132874 $ &$ 0.192060 $ &$ 0.243668 $ &$ 0.567879 $\\
\hline
\end{tabular}
\label{tab:QNMs}
\caption{Scalar QNM frequencies for a selection of modes at different spins. We provide both the value $\omega^{(0)}$ in GR as well as the cubic even correction $\omega^{(1)}$. For each frequency we provide the first six digits, which are significant for all modes.}
\end{table*}

\newpage

\section{Supplementary error estimates}
To show the dependence of the convergence on the multipole $l$, we include Fig.~\ref{fig:accuracy_l012345} that depicts the estimated relative error on the fundamental co-rotating modes $l=m$. Note that the higher the multipole number $l$, the better the numerical results converge. 

\begin{figure}[h]
    \centering
    \subfloat[$0\le l \le 2$, $m=l$, $n=0$]{
    \includegraphics[width=0.48\linewidth]{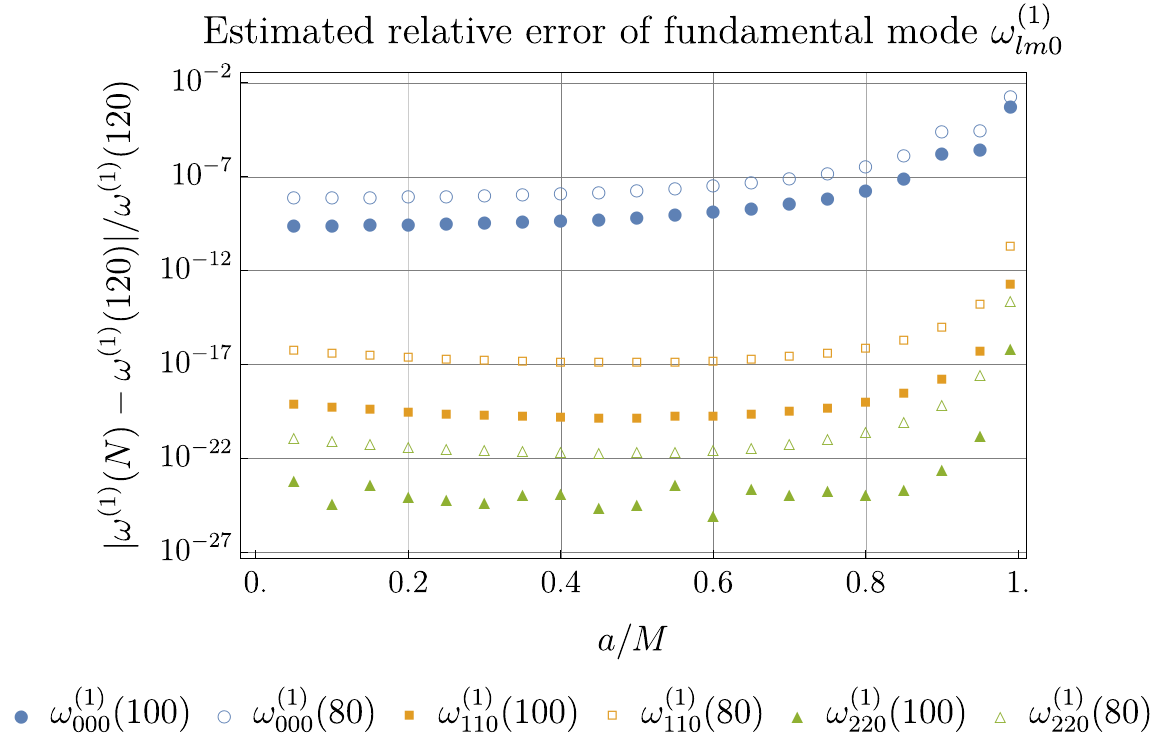}
    \label{fig:accuracy_n0l012}
    }%
    \hfill
    \subfloat[$3\le l \le 5$, $m=l$, $n=0$]{%
    \includegraphics[width=0.48\linewidth]{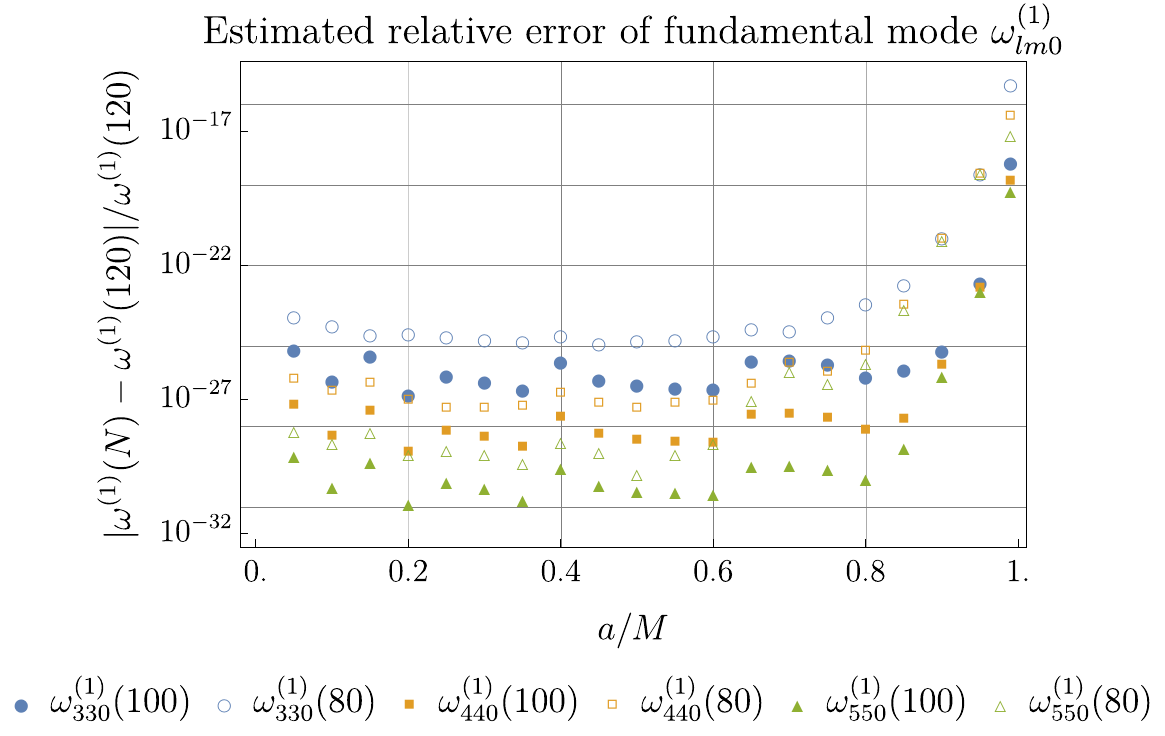}
    \label{fig:accuracy_n0l345}
    }%
    \caption{Estimated relative errors for corrections $\omega^{(1)}$ to the fundamental modes with with $0\le l \le 5$ and $l=m$ as a function of spin. Filled markers indicate the relative errors between high resolution $N_r=N_y=120$ and low  resolution $N_r=N_y=80$, and filled markers between the same high resolution and moderate resolution $N_r=N_y=100$. To prevent clutter the plot is split up in the modes with $0\le l \le 2$ and  $3\le l \le 5$}
    \label{fig:accuracy_l012345}
\end{figure}

\end{document}